\documentclass[12pt]{article}                  %% LaTeX 2e
\usepackage{osajnl}
\usepackage{overcite,hyperref}

\begin{document}
\def\baselinestretch{1.0}

\title{High-precision Absolute Distance and Vibration Measurement 
using Frequency Scanned Interferometry}

\author{Hai-Jun Yang$^*$, Jason Deibel, Sven Nyberg, Keith Riles}

%% for REVTeX4, each author name can be set in a separate \author{} field

\affiliation{Department of Physics, University of Michigan, Ann Arbor, MI 48109-1120, USA}

\begin{abstract}
In this paper, we report high-precision absolute distance and vibration
measurements performed with frequency scanned interferometry 
using a pair of single-mode optical fibers. Absolute distance was determined 
by counting the interference fringes produced while scanning the laser frequency. 
A high-finesse Fabry-Perot interferometer(F-P) was used to determine frequency 
changes during scanning. Two multiple-distance-measurement analysis techniques 
were developed to improve distance precision and to extract the amplitude
and frequency of vibrations.
Under laboratory conditions, measurement precision of $\sim$ 50 nm was achieved for 
absolute distances ranging from 0.1 meters to 0.7 meters by using the first
multiple-distance-measurement technique. 
The second analysis technique has the capability to 
measure vibration frequencies ranging from 0.1 Hz to 100 Hz with amplitude
as small as a few nanometers, without a {\em priori} knowledge.
\end{abstract}

% instrumentation, fringe analysis, interferometry, vibration analysis

\maketitle %% NULL FUNCTION WITH LATEX 2e

%\twocolumn

\section{Introduction}

The motivation for this project is to design a novel optical
system for quasi-real time alignment of tracker detector elements
used in High Energy Physics (HEP) experiments. A.F. Fox-Murphy {\em et.al.}
from Oxford University reported their design of a frequency
scanned interferometer (FSI) for precise alignment of the ATLAS
Inner Detector \cite{ox98}. Given the demonstrated need for
improvements in detector performance, we plan to design an
enhanced FSI system to be used for the alignment of tracker
elements in the next generation of electron positron Linear
Collider (ILC) detectors. Current plans for future detectors
require a spatial resolution for signals from a tracker detector, 
such as a silicon microstrip or silicon drift detector,
to be approximately 7-10 $\mu m$\cite{orangebook}. To achieve this
required spatial resolution, the measurement precision of absolute
distance changes of tracker elements in one dimension should be on the order of 1
$\mu m$. Simultaneous measurements from hundreds of interferometers
will be used to determine the 3-dimensional positions of the
tracker elements.

We describe here a demonstration
FSI system built in the laboratory for initial feasibility studies.  
The main goal was to determine the potential accuracy of absolute distance 
measurements (ADM's) that could be achieved under controlled conditions.  
Secondary goals included estimating the effects of vibrations and studying 
error sources crucial to the absolute distance accuracy.  A significant 
amount of research on ADM's using wavelength scanning interferometers already 
exists \cite{stone99,dai98,barwood98,bechstein98,thiel95,kikuta86}.
In one of the most comprehensive publications on this subject,
Stone {\em et al.} describe in detail a wavelength scanning heterodyne interferometer
consisting of a system built around both a reference and a measurement
interferometer\cite{stone99}, the measurement precisions of absolute distance
ranging from 0.3 to 5 meters are $\sim$ 250 nm by averaging distance
measurements from 80 independent scans.

Detectors for HEP experiment must usually be operated remotely for 
safety reasons because of intensive radiation, high voltage or strong magnetic
fields. In addition, precise tracking elements are typically surrounded by
other detector components, making access difficult. For practical HEP application
of FSI, optical fibers for light delivery and return are therefore necessary.

We constructed a FSI demonstration system by employing a pair of single-mode 
optical fibers of approximately 1 meter length each, one for transporting
the laser beam to the beam splitter and retroreflector and another for 
receiving return beams. A key issue for the optical fiber FSI is that the 
intensity of the return beams received by the optical fiber is very weak;
the natural geometrical efficiency is $~ 6.25 \times 10^{-10}$ for a measurement
distance of 0.5 meter. In our design, we use a gradient index 
lens (GRIN lens) to collimate the output beam from the optical fiber.

We believe our work represents a significant advancement in the field of
FSI in that high-precision ADM's and vibration measurements are performed 
(without a {\em priori} knowledge of vibration strengths and frequencies), 
using a tunable laser, an isolator, an off-the-shelf F-P, a fiber coupler, 
two single-mode optical fibers, an interferometer and novel fringe analysis and 
vibration extraction techniques. Two new multiple-distance-measurement analysis 
techniques are presented, to improve precision and to extract the amplitude 
and frequency of vibrations. 
Expected dispersion effects when a corner cube prism or a beamsplitter
substrate lies in the interferometer beam path are confirmed,
and observed results agree well with
results from numerical simulation. When present, the dispersion effect has 
a significant impact on the absolute distance measurement. 
The limitations of our current FSI system are also discussed in the paper, and
major uncertainties are estimated.

\section{Principles}

The intensity $I$ of any two-beam interferometer can be expressed as
$$
\begin{array}{c}
I = I_1 + I_2 + 2\sqrt{I_1 I_2} \cos(\phi_1 - \phi_2)
\end{array}
\eqno{(1)}
$$
where $I_1$ and $I_2$ are the intensities of the two combined beams,
$\phi_1$ and $\phi_2$ are the phases.
Assuming the optical path lengths of the two beams are $L_1$ and
$L_2$, the phase difference in Eq.~(1) is
$\Phi = \phi_1 - \phi_2 = 2\pi |L_1 - L_2|(\nu/c)$,
where $\nu$ is the optical frequency of the laser beam, and c is
the speed of light.

For a fixed path interferometer, as the frequency of the laser is
continuously scanned, the optical beams will
constructively and destructively interfere, causing ``fringes''.
The number of fringes $\Delta N$ is
$$
\begin{array}{c}
\Delta N = |L_1 - L_2|(\Delta\nu/c) = L\Delta\nu/c
\end{array}
\eqno{(2)}
$$
where $L$ is the optical path difference between the two beams,
and $\Delta\nu$ is the scanned frequency range. The optical path
difference (OPD for absolute distance between beamsplitter and retroreflector)
can be determined by counting interference fringes while scanning the laser frequency. 

\begin{figure}[htbp]
\centerline{\scalebox{0.5}{\includegraphics{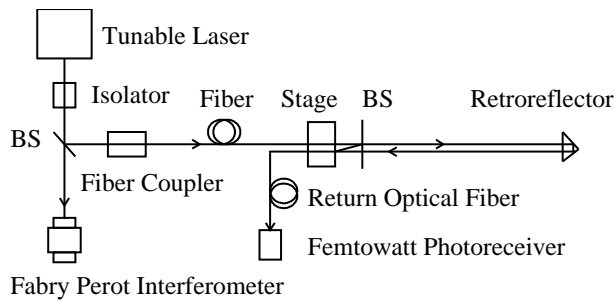}}}
\caption{Schematic of an optical fiber FSI system.}
\end{figure}

\section{Demonstration System of FSI}

A schematic of the FSI system with a pair of optical fibers is shown in Fig.1.  
The light source is a New Focus Velocity 6308 tunable laser 
(665.1 nm $<\lambda<$ 675.2 nm). A high-finesse ($>200$) Thorlabs SA200
F-P is used to measure the frequency range scanned by the laser. The free 
spectral range (FSR) of two adjacent F-P peaks is 1.5 GHz, which corresponds 
to 0.002 nm. A Faraday Isolator was used to reject light reflected back into
the lasing cavity. The laser beam was coupled into a single-mode optical fiber
with a fiber coupler.
Data acquisition is based on a National Instruments DAQ
card capable of simultaneously sampling 4 channels at a rate of 5 MS/s/ch with
a precision of 12-bits.  Omega thermistors with a tolerance of 0.02 K and a
precision of 0.01 $mK$ are used to monitor temperature.  The apparatus
is supported on a damped Newport optical table.

In order to reduce air flow and temperature fluctuations, a transparent plastic
box was constructed on top of the optical table. PVC pipes were installed to 
shield the volume of air surrounding the laser beam. Inside the PVC pipes, 
the typical standard deviation of 20 temperature measurements was about $0.5 ~mK$. 
Temperature fluctuations were suppressed by a factor of approximately 100 by 
employing the plastic box and PVC pipes.

The beam intensity coupled into the return optical fiber is very weak, requiring
ultra-sensitive photodetectors for detection. Considering the limited laser beam 
intensity and the need to split into many beams to serve a set
of interferometers, it is vital to increase the geometrical efficiency.
To this end, a collimator is built by placing an optical fiber in a ferrule 
(1mm diameter) and gluing one end of the optical fiber to a GRIN lens. 
The GRIN lens is a 0.25 pitch lens with  0.46 numerical aperture, 1 mm diameter 
and 2.58 mm length which is optimized for a wavelength of 630nm. 
The density of the outgoing beam from the optical 
fiber is increased by a factor of approximately 1000 by using a GRIN lens. 
The return beams are received by another optical fiber and amplified by a 
Si femtowatt photoreceiver with a gain of $2 \times 10^{10} V/A$.

\section{Multiple-Distance-Measurement Techniques}

For a FSI system, drifts and vibrations occurring along the optical
path during the scan will be magnified by a factor of $\Omega =
\nu / \Delta \nu$, where $\nu$ is the average optical frequency of the laser
beam and $\Delta \nu$ is the scanned frequency. For the full scan of
our laser, $\Omega \sim 67 $. Small vibrations and
drift errors that have negligible effects for many optical
applications may have a significant impact on a FSI system.
A single-frequency vibration
may be expressed as $x_{vib}(t) = a_{vib} \cos(2\pi f_{vib} t + \phi_{vib})$,
where $a_{vib}$, $f_{vib}$ and $\phi_{vib}$ are the amplitude, frequency and phase of the
vibration respectively.
If $t_0$ is the start time of the scan, Eq.~(2) can be re-written as
$$
\begin{array}{c}
\Delta N=L\Delta\nu/c+2[x_{vib}(t)\nu(t)-x_{vib}(t_0)\nu(t_0)]/c
\end{array}
\eqno{(3)}
$$
If we approximate $\nu(t) \sim \nu(t_0) = \nu$, the
measured optical path difference $L_{meas}$ may be expressed as
$$
\begin{array}{c}
L_{meas} = L_{true} - 4 a_{vib} \Omega \sin[\pi f_{vib}(t-t_0)] \times \\
\sin[\pi f_{vib}(t+t_0)+\phi_{vib}]
\end{array}
\eqno{(4)}
$$
where $L_{true}$ is the true optical path difference without
vibration effects. If the path-averaged refractive index of
ambient air $\bar{n}_g$ is known, the measured distance
is $R_{meas} = L_{meas}/(2\bar{n}_g)$.

If the measurement window size $(t-t_0)$ is fixed and the window used to
measure a set of $R_{meas}$ is sequentially shifted, the
effects of the vibration will be evident. 
We use a set of distance measurements in one scan by successively shifting the 
fixed-length measurement window one F-P peak forward each time. 
The arithmetic average of all measured $R_{meas}$ values in one scan is taken to be the
measured distance of the scan (although more sophisticated fitting methods
can be used to extract the central value). 
For a large number of
distance measurements $N_{meas}$, the vibration effects can be greatly suppressed. 
Of course, statistical uncertainties
from fringe and frequency determination, dominant in
our current system, can also be reduced with multiple scans.
Averaging multiple measurements in one scan, however, provides similar precision
improvement as averaging distance measurements from 
independent scans, and is faster, more efficient, and less 
susceptible to systematic errors from drift.
In this way, we can improve the distance accuracy dramatically if 
there are no significant drift errors during one scan, caused, for example, 
by temperature variation. 
This multiple-distance-measurement technique is called 'slip measurement
window with fixed size', shown in Fig.2. However, there is a trade off in that the
thermal drift error is increased with the increase of $N_{meas}$
because of the larger magnification factor $\Omega$ for a smaller
measurement window size.

\begin{figure}[htbp]
\centerline{\scalebox{0.35}{\includegraphics{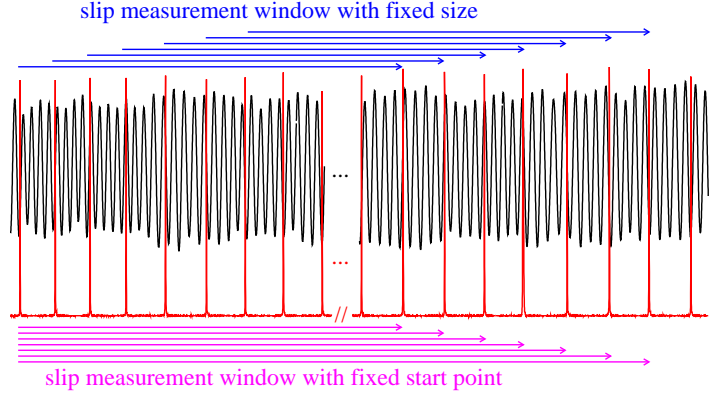}}}
\caption{The schematic of two multiple-distance-measurement techniques.
The interference fringes from the femtowatt photoreceiver and
the scanning frequency peaks from the Fabry-Perot interferometer(F-P)
for the optical fiber FSI system recorded simultaneously by DAQ card
are shown in black and red, respectively. The free spectral range(FSR)
of two adjacent F-P peaks (1.5 GHz) provides a calibration of the scanned frequency range.}
\end{figure}

In order to extract the amplitude and frequency of the vibration, another 
multiple-distance-measurement technique called 'slip measurement window with 
fixed start point' is shown in Fig.2. In Eq.~(3), if $t_0$ is fixed, the 
measurement window size is enlarged one F-P peak for each shift, an 
oscillation of a set of measured $R_{meas}$ values reflects the amplitude 
and frequency of vibration. 
This technique is not suitable for distance measurement because there
always exists an initial bias term including $t_0$
which cannot be determined accurately in our current system.

\section{Absolute Distance Measurement}

The typical measurement residual versus the distance
measurement number in one scan using the above technique is shown in Fig.3(a), where
the scanning rate was 0.5 nm/s and the sampling rate was 125 kS/s.
Measured distances minus their average value for 10 sequential scans
are plotted versus number of measurements ($N_{meas}$) per scan in Fig.3(b).
The standard deviations (RMS) of distance measurements for 
10 sequential scans are plotted versus number of measurements ($N_{meas}$) per scan in Fig.3(c).
It can be seen that the distance errors decrease with an increase of $N_{meas}$.
The RMS of measured distances for 10 sequential scans
is 1.6 $\mu m$ if there is only one distance measurement per scan ($N_{meas}=1$).
If $N_{meas}=1200$ and the average value of 1200 distance measurements
in each scan is considered as the final measured distance of the scan,
the RMS of the final measured distances for 10 scans is 41 nm for the distance of
449828.965 $\mu m$, the relative distance measurement precision is 91 ppb.

\begin{figure}[htbp]
\centerline{\scalebox{0.6}{\includegraphics{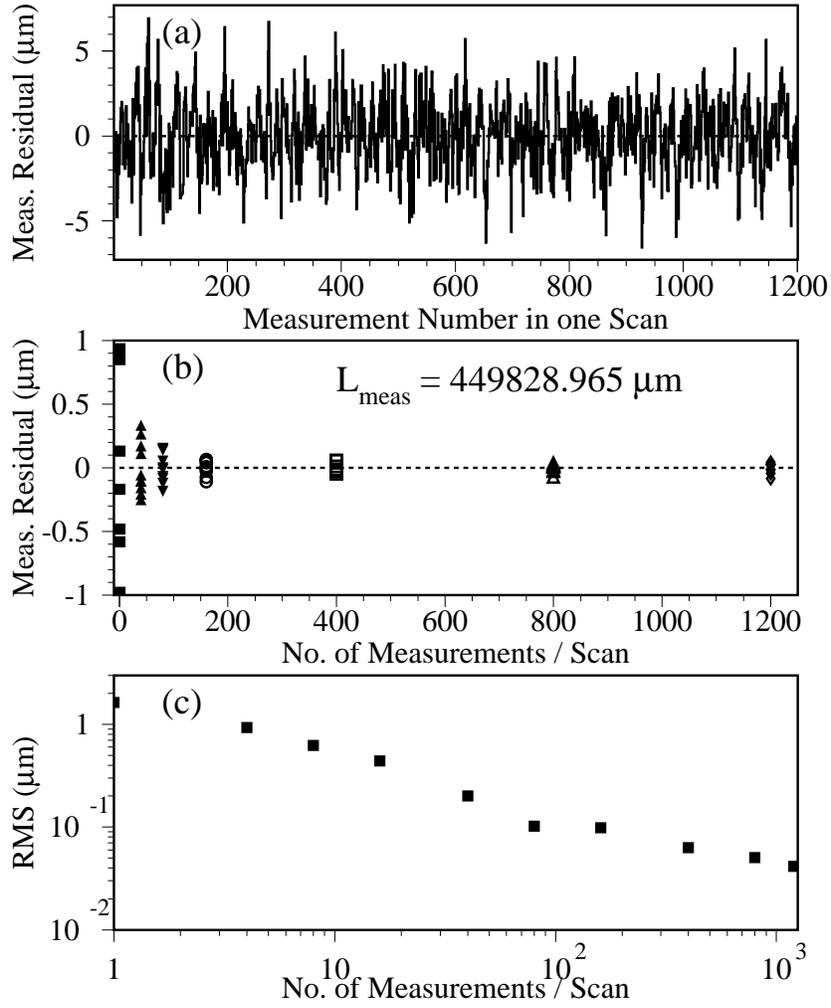}}}
\caption{Distance measurement residual spreads versus number of distance measurement 
$N_{meas}$ 
(a) for one typical scan, 
(b) for 10 sequential scans,
(c) is the standard deviation of distance measurements for 10 sequential scans
versus $N_{meas}$.}
\end{figure}

\begin{figure}[htbp]
\centerline{\scalebox{0.6}{\includegraphics{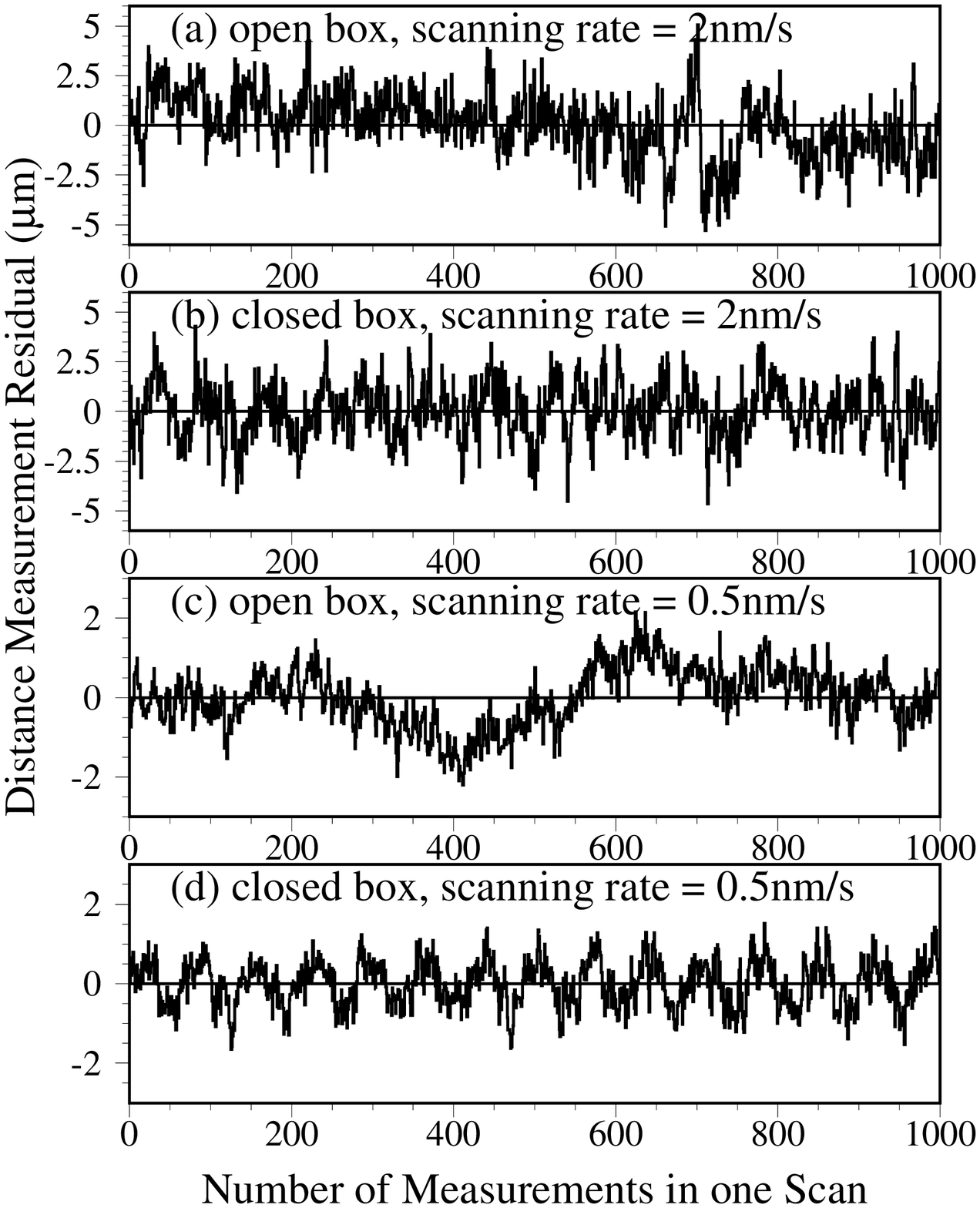}}}
\caption{Distance measurement residual spreads versus $N_{meas}$ in one scan;
(a) for the open box with a scanning rate of 2 nm/s, 
(b) for the closed box with a scanning rate of 2 nm/s, 
(c) for the open box with a scanning rate of 0.5 nm/s, 
(d) for the closed box with a scanning rate of 0.5 nm/s.}
\end{figure}

Some typical measurement residuals are plotted versus the number of distance measurements
in one scan($N_{meas}$) for open box and closed box data with scanning 
rates of 2 nm/s and 0.5 nm/s in Fig.4(a,b,c,d), respectively. 
The measured distance is approximately 10.4 cm.
It can be seen that the slow fluctuations of multiple distance measurements
for open box data are larger than that for closed box data.

The standard deviation (RMS) of measured distances for 10 sequential scans 
is approximately 1.5 $\mu m$ if there is only one distance measurement per scan
for closed box data. By using multiple-distance-measurement 
technique, the distance measurement precisions for various closed box data
with distances ranging from 10 cm to 70 cm collected in the past year
are improved significantly, precisions of approximately 50 nanometers are
demonstrated under laboratory conditions, as shown in Table 1. 
All measured precisions
listed in the Table 1. are the RMS of measured distances for 10 sequential scans.
Two FSI demonstration systems, 'air FSI' and 'optical fiber FSI',
are constructed for extensive tests of multiple-distance-measurement technique, 
'air FSI' means FSI with the laser beam transported entirely in the ambient atmosphere,
'optical fiber FSI' represents FSI with the laser beam delivered to the interferometer 
and received back by single-mode optical fibers. 

\begin{table}
\begin{tabular}{|c|c|c|c|c|} \hline
Distance & \multicolumn{2}{|c|}{Precision($\mu m$)} & Scanning Rate & FSI System \\ 
\cline{2-3}
(cm) & open box & closed box & (nm/s) & (Optical Fiber or Air) \\ \hline
10.385107 & 1.1 & 0.019 & 2.0 & Optical Fiber FSI \\ \hline
10.385105 & 1.0 & 0.035 & 0.5 & Optical Fiber FSI \\ \hline
20.555075 & - & 0.036, 0.032 & 0.8 & Optical Fiber FSI \\ \hline
20.555071 & - & 0.045, 0.028 & 0.4 & Optical Fiber FSI \\ \hline
41.025870 & 4.4 & 0.056, 0.053 & 0.4 & Optical Fiber FSI \\ \hline
44.982897 & - & 0.041 & 0.5 & Optical Fiber FSI \\ \hline 
61.405952 & - & 0.051 & 0.25 & Optical Fiber FSI \\ \hline
65.557072 & 3.9, 4.7 & - & 0.5 & Air FSI \\ \hline
70.645160 & - & 0.030, 0.034, 0.047 & 0.5 & Air FSI \\ \hline
\end{tabular}
\caption{Distance measurement precisions for various setups using the 
multiple-distance-measurement
technique.}
\end{table}

Based on our studies, the slow fluctuations are reduced to a negligible level by using 
the plastic box and PVC pipes to suppress temperature fluctuations.
The dominant error comes from the uncertainties of the interference fringes 
number determination; the fringes uncertainties are uncorrelated for multiple 
distance measurements. In this case, averaging multiple distance measurements
in one scan provides a similar precision improvement to averaging distance measurements
from multiple independent scans, but is faster, more efficient and less susceptible
to systematic errors from drift. But, for open box data, the slow fluctuations 
are dominant, on the order of few microns in our laboratory. The 
measurement precisions for single and multiple distance open-box measurements 
are comparable, 
which indicates that the slow fluctuations cannot be adequately suppressed by using
the multiple-distance-measurement technique.
A dual-laser FSI system\cite{bechstein98,coe2001} intended to cancel the drift 
error is currently under study in our laboratory 
(to be described in a subsequent article).

From Fig.4(d), we observe periodic oscillation of the
distance measurement residuals in one scan, the fitted frequency
is $3.22 \pm 0.01$ Hz for the scan. The frequency depends on the 
scanning rate, $f \sim (scanning~~rate~~ in~~ nm/s) \times  60/(675.1 nm - 665.1 nm)$.
From Eq.(4), it is clear that the amplitude of the vibration or
oscillation pattern for multiple distance measurements 
depends on $4 a_{vib} \Omega \sin[\pi f_{vib}(t-t_0)] $.
If $a_{vib}$, $f_{vib}$ are constant values, it depends on the size of 
the distance measurement window. 
Subsequent investigation with a CCD camera trained on
the laser output revealed that the apparent 
$\sim$3 Hz vibration during the 0.5 nm/s scan arose from the beam's centroid motion.
Because the centroid motion is highly reproducible,
we believe that the effect comes from motion of the internal 
hinged mirror in the laser used to scan its frequency.

The measurable distance range is limited in our current optical
fiber FSI demonstration system for several reasons. 
For a given scanning rate of 0.25 nm/s,
the produced interference fringes, estimated by
$\Delta N \sim (2 \times \Delta L \times \Delta \nu)/c$,
are approximately 26400 in a 40-second scan for a measured distance 
($\Delta L$) of 60 cm, that is $\sim 660 $ fringes/s, where $\Delta \nu$ is 
the scanned frequency and $c$ is the speed
of light. The currently used femtowatt photoreceiver has 3 dB frequency
bandwidth ranging from 30-750 Hz, the transimpedance gain decreases quickly beyond 750 Hz.
There are two ways to extend the measurable distance range. One straightforward way is to
extend the effective frequency bandwidth of the femtowatt photoreceiver; 
the other way is to decrease the interference fringe rate by decreasing the laser scanning rate. 
There are two major drawbacks for the second way; one is that larger
slow fluctuations occur during longer scanning times; the other is that the laser scanning
is not stable enough to produce reliable interference fringes if the scanning rate 
is lower than 0.25 nm/s for our present tunable laser. In addition, another 
limitation to distance range is that the intensity of the return beam from the 
retroreflector decreases inverse-quadratically with range.

\section{Vibration Measurement}

In order to test the vibration measurement technique, 
a piezoelectric transducer (PZT) was employed to produce 
vibrations of the retroreflector. For instance, the frequency of the controlled 
vibration source was set to $1.01 \pm 0.01$ Hz with amplitude of $0.14 \pm 0.02 ~\mu m$.
For $N_{meas} =2000$ distance measurements in one scan, the magnification factor for each
distance measurement depends on the scanned frequency of the measurement window,
$\Omega(i) = \nu/ \Delta \nu(i)$, where, $\nu$ is the average frequency of the laser beam
in the measurement window, scanned frequency $\Delta \nu(i) = (4402-N_{meas}+i)\times 1.5$ GHz,
where i runs from 1 to $N_{meas}$, shown in Fig.5(a). The distance measurement residuals for
2000  distance measurements in the scan are shown in Fig.5(b), the oscillation of the 
measurement residuals reflect the vibration of the retroreflector.
Since the vibration is magnified by a factor of $\Omega(i)$ for each distance measurement, 
the corrected measurement residuals are measurement residuals divided by the corresponding
magnification factors, shown in Fig.5(c).
The extracted vibration frequencies and amplitudes using this technique
are $f_{vib} = 1.007 \pm 0.0001$~Hz, $A_{vib} = 0.138 \pm 0.0003 ~\mu m$,
respectively, in good agreement with expectations.

\begin{figure}[htbp]
\centerline{\scalebox{0.6}{\includegraphics{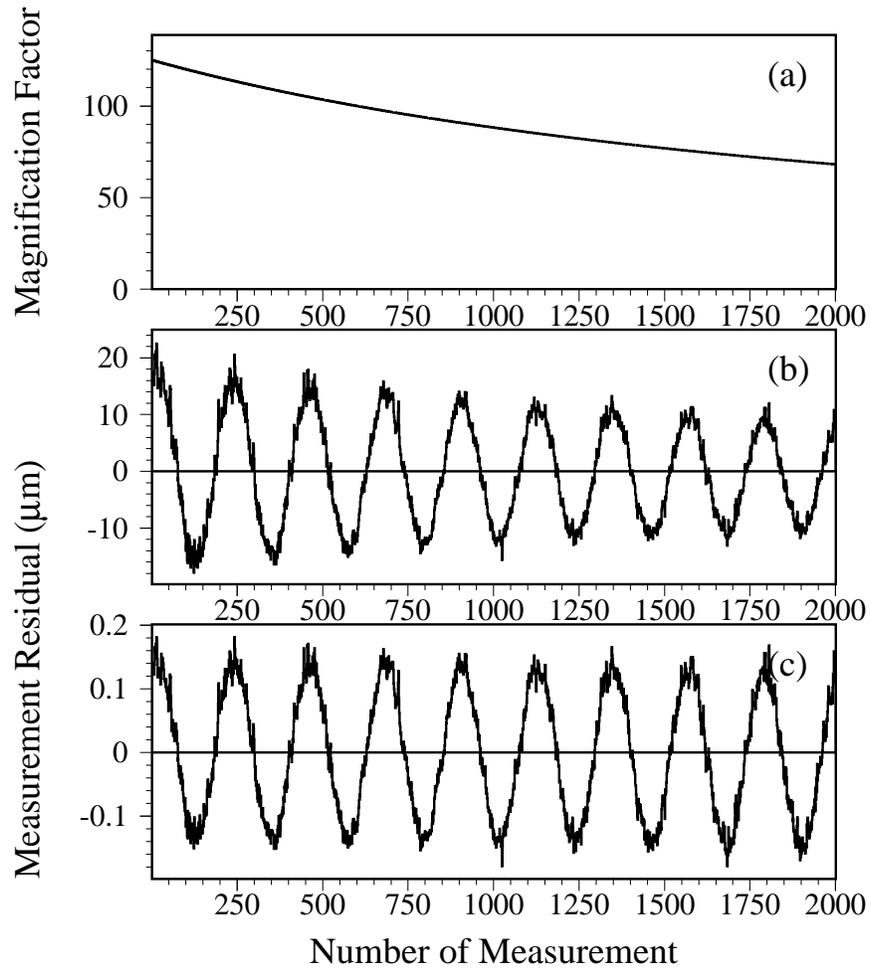}}}
\caption{The frequency and amplitude of the controlled vibration source 
are 1 Hz and 140 nanometers, 
(a) Magnification factor  versus number of distance measurements,
(b) Distance measurement residual  versus number of distance measurements,
(c) Corrected measurement residual  versus number of distance measurements.}
\end{figure}

\begin{figure}[htbp]
\centerline{\scalebox{0.6}{\includegraphics{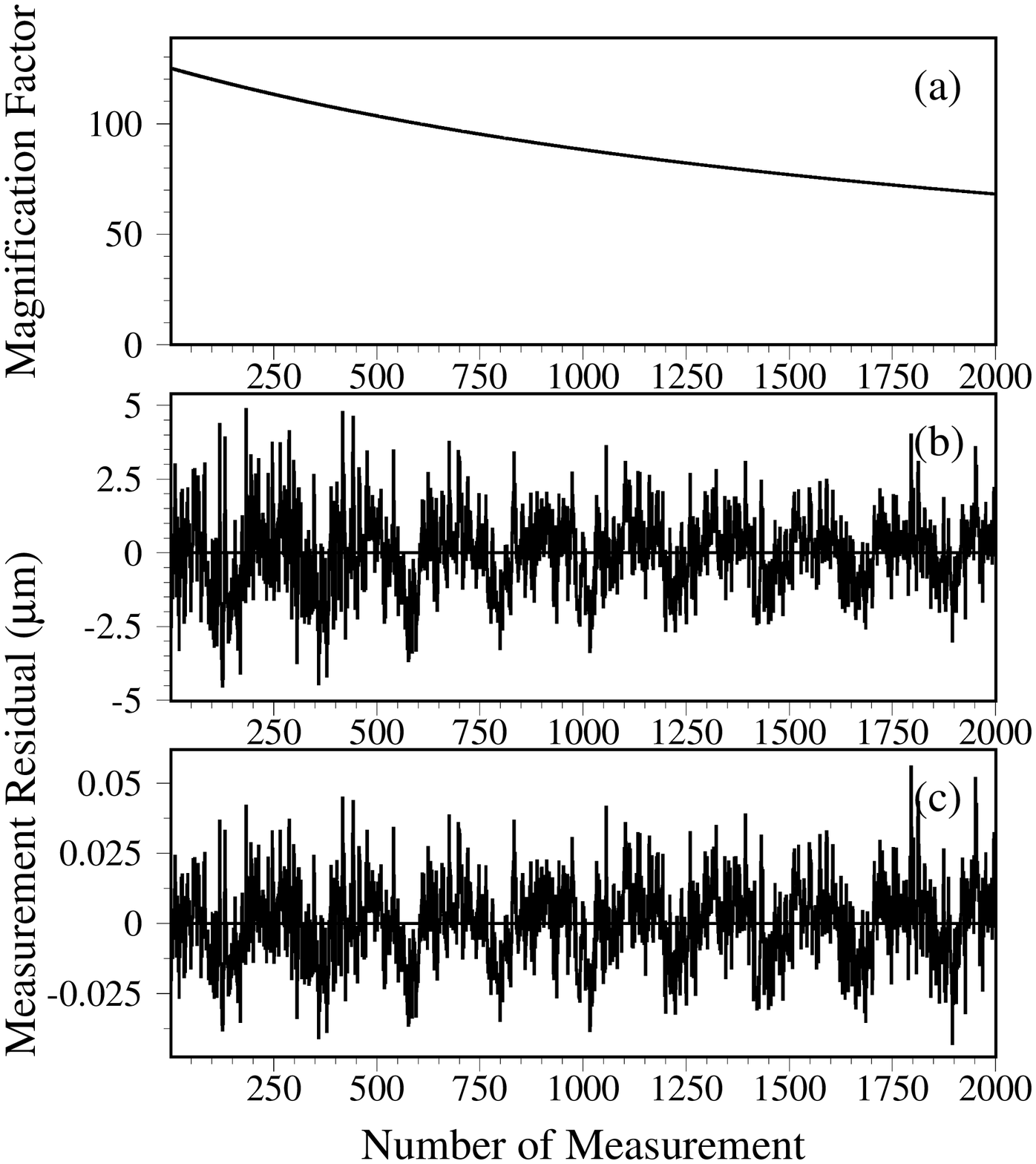}}}
\caption{The frequency and amplitude of the controlled vibration source 
are 1 Hz and 9.5 nanometers, 
(a) Magnification factor  versus number of distance measurements,
(b) Distance measurement residual  versus number of distance measurements,
(c) Corrected measurement residual  versus number of distance measurements.}
\end{figure}

Another demonstration was made for the same vibration frequency, but with
an amplitude of only $9.5 \pm 1.5$ nanometers. The magnification factors,
distance measurement residuals and corrected measurement residuals for 2000
measurements in one scan are shown in Fig.6(a), Fig.6(b) and Fig.6(c), respectively.
The extracted vibration frequencies and amplitudes using this technique
are $f_{vib} = 1.025 \pm 0.002$~Hz, $A_{vib} = 9.3 \pm 0.3 $ nanometers.

In addition, vibration frequencies at 
0.1, 0.5, 1.0, 5, 10, 20, 50, 100 Hz with controlled vibration amplitudes ranging from 
9.5 nanometers to 400 nanometers were studied extensively using our current
FSI system. The measured vibrations and expected vibrations all 
agree well within the 10-15\% level for amplitudes, 1-2\% for frequencies,
where we are limited by uncertainties in the expectations.
Vibration frequencies far below 0.1 Hz can be regarded as slow fluctuations, 
which can not be suppressed by the above analysis techniques. 

For comparison, nanometer vibration measurement by a self-aligned optical feedback vibrometry
technique has been reported\cite{otsuka2002}.
The vibrometry technique is able to measure vibration frequencies ranging from 
20 Hz to 20 kHz with minimal measurable vibration amplitude of 1 nm. 
Our second multiple-distance-measurement technique demonstrated above has capability to 
measure vibration frequencies ranging from 0.1 Hz to 100 Hz with minimal amplitude
on the level of several nanometers, without a {\em priori} knowledge of the
vibration strengths or frequencies.

\section{Impact of Dispersion Effects on Distance Measurement}

Dispersive elements such as a beamsplitter, corner
cube prism, etc. in the interferometer 
can create an apparent offset in measured distance for an FSI system,
since the optical path length of the dispersive element changes during the scan.
The small OPD change caused by dispersion 
is magnified by a factor of $\Omega$ and has a significant 
effect on the absolute distance measurement for the FSI system.
The measured optical path difference $L_{meas}$ may be expressed as
$$
\begin{array}{rl}
L_{meas} & = | L(t)/\lambda(t) - L(t0)/\lambda(t0) | 
             \times c/\Delta \nu  \\
L(t) & = 2 \times ( D1 \times n_{air} + D2 \times n(\lambda(t))_{cornercube} )
\end{array}
\eqno{(5)}
$$
where $L(t)$ and $L(t0)$ refer to the OPD at times t and t0, respectively,
$\lambda(t)$ and $\lambda(t0)$ are the wavelength of the laser beam at times t
and t0, c is the speed of light, D1 and D2 are true geometrical distances in the air
and in the corner cube prism, $n_{air}$ and $n(\lambda(t))_{cornercube}$ are the refractive
index of ambient atmosphere and the refractive index of the corner cube prism 
for $\lambda(t)$, respectively. The measured distance $R_{meas} = L_{meas}/(2\bar{n}_g)$,
where $\bar{n}_g$ is the average refractive index around the optical path.

The Sellmeier formula for dispersion in crown glass (BK7)\cite{schott96} can be written as,
$$
n^2(\lambda) = 1 + \frac{B_1\lambda^2}{\lambda^2-C_1} + \frac{B_2\lambda^2}{\lambda^2-C_2}
+ \frac{B_3\lambda^2}{\lambda^2-C_3}
\eqno{(6)}
$$
where, the beam wavelength $\lambda$ is in unit of microns, 
$B_1 = 1.03961212$, $B_2 = 0.231792344$, $B_3 = 1.01046945$,
$C_1 = 0.00600069867$, $C_2 = 0.0200179144$, $C_3 = 103.560653$.

If we use the first multiple-distance-measurement technique described above 
to make 2000 distance measurements for one typical scan, where 
the corner cube prism is used as retroreflector, 
we observe a highly reproducible drift 
in measured distance, as shown in Fig.7, where the fitted distance drift 
is $6.14 \pm 0.08$ microns for one typical scan using a straight line fit. 
However, there is no apparent drift if we replace the corner cube 
prism by the hollow retroreflector.

\begin{figure}[htbp]
\centerline{\scalebox{0.5}{\includegraphics{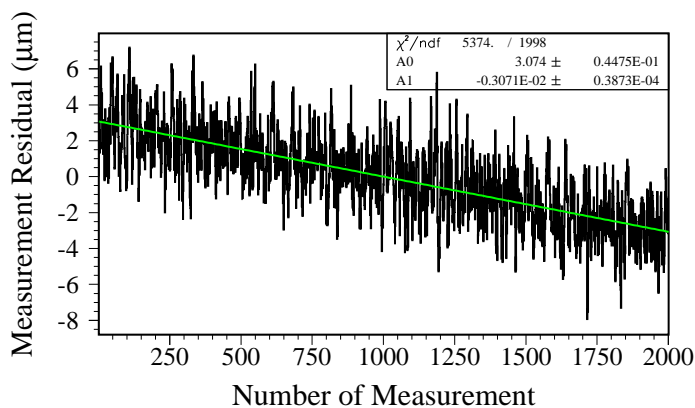}}}
\caption{Residuals of 2000 distance measurements for one typical scan, the corner cube
prism is used as retroreflector.}
\end{figure}

Numerical simulations have been carried out using Eq.(5) and Eq.(6) to understand the above phenomena. 
For instance, consider the case D1 = 20.97 cm and D2 = 1.86 cm (the uncertainty of D2 is 0.06 cm), 
where the first and the last measured distances among 2000 sequential
distance measurements are denoted $R_1$ and $R_{2000}$, respectively. 
Using the Sellmeier equation (Eq.(6)) for modeling the corner cube 
prism material (BK7) dispersion, 
we expect $R_1 - (D1 + D2)$ = 373.876 microns and $R_{2000} - (D1 + D2)$ = 367.707 microns.
The difference between $R_1$ and $R_{2000}$ is $6.2 \pm 0.2$ microns which 
agrees well with our observed $6.14 \pm 0.08$ microns drift over 2000 measured distances. The
measured distance shift and drift strongly depend on D2, but
are insensitive to D1. A change of 1 cm in D1 leads to a 3-nanometer distance shift, 
but the same change in D2 leads to a 200-micron distance shift.
If a beamsplitter is oriented with its reflecting side facing the laser beam,
then there is an additional dispersive distance shift. We have verified this effect
with 1-mm and a 5-mm beam splitters. 
When we insert an additional
beamsplitter with 1 mm thickness between the retroreflector and the
original beamsplitter in the optical fiber FSI system, we observe
a 500 microns shift on measured distance if $\bar{n}_g$ is fixed 
consistent with the numerical simulation result. 
For the 5-mm beam splitter (the measured thickness of the beam splitter is $4.6 \pm 0.05$ mm)
, the first 20 scans were performed with the beamsplitter's
anti-reflecting surface facing the optical fibers and the second 20 scans
with the reflecting surface facing the optical fibers. The expected 
drifts ($R_{2000} - R_{1}$) for the first and the second 20 scans from the dispersion effect are 
0 and $-1.53 \pm 0.05$ microns, respectively. The measured drifts by averaging measurements from 
20 sequential scans are $-0.003 \pm 0.12$ microns and $-1.35 \pm 0.17$ microns, respectively. 
The measured values agree well with expectations. 
In addition, the dispersion effect from air\cite{peck72,ciddor96} is also estimated by using
numerical simulation. The expected drift ($R_{2000} - R_{1}$) from air dispersion
is approximately -0.07 microns for an optical path of 50 cm in air, this effect cannot be
detected for our current FSI system. However, it could be verified by using a FSI with a vacuum 
tube surrounding the laser beam; the measured distance with air in the tube would be approximately 4 
microns larger than for an evacuated tube.

%%%

In summary, dispersion effects can have a significant impact on absolute
distance measurements, but can be minimized with care for elements placed in 
the interferometer or corrected for, once any necessary dispersive elements 
in the interferometer are understood.

\section{Error Estimations}

Some major error sources are estimated in the following;

1) Error from uncertainties of fringe and scanned frequency determination. 
The measurement precision of the distance $R$ 
(the error due to the air's refractive index uncertainty 
is considered separately below) is given by
$
 (\sigma_R/R)^2 = (\sigma_{\Delta N}/{\Delta N})^2 + (\sigma_{\Delta \nu}/{\Delta \nu})^2
$.
where $R$, $\Delta N$, $\Delta \nu$, $\sigma_R$, $\sigma_{\Delta N}$, $\sigma_{\Delta \nu}$ 
are measurement distance, fringe numbers, scanned frequency and their corresponding
errors. For a typical scanning rate of 0.5 nm/s with a 10 nm scan range, the
full scan time is 20 seconds. The total number of samples for one
scan is 2.5 MS at a sampling rate of 125 kS/s. There is about a 
4$\sim$5 sample ambiguity in fringe peak and valley position due
to a vanishing slope and the limitation of the 12-bit sampling
precision. However, there is a much smaller uncertainty for the F-P
peaks because of their sharpness. Thus, the
estimated uncertainty is $\sigma_R/R \sim 1.9 ~ppm$ for one full scan for a 
magnification factor $\Omega = 67$.
If the number of distance measurements $N_{meas} = 1200$, 
the distance measurement window is smaller, the corresponding magnification factor is 
$\Omega^* = \nu/ \Delta \nu$, where, $\nu$ is the average frequency of the laser beam, 
$\Delta \nu = (4402-N_{meas})\times 1.5$ GHz.
One obtains $\Omega^* \sim 94$, $\sigma_R/R
\sim 1.9 ~ppm \times \Omega^*/ \Omega /\sqrt{N_{meas}} \sim 77 ~ppb$.  

2) Error from vibrations. The detected amplitude and frequency for
vibration (without controlled vibration source) are about 0.3 $\mu m$ 
and 3.2 Hz. The corresponding time for
$N_{meas} = 1200$ sequential distance measurements is 5.3 seconds. 
A rough estimation of the resulting error gives $\sigma_R/R \sim 
0.3~\mu m /(5.3~s\times 3.2~Hz \times 4)/R \sim 10 ~ppb$
for a given measured distance $R=0.45$ meters.

3) Error from thermal drift. The
refractive index of air depends on air temperature, humidity and
pressure (fluctuations of humidity and pressure have
negligible effects on distance measurements for the 20-second scan). 
Temperature fluctuations are well controlled down to
about $0.5 ~mK$(RMS) in our laboratory by the plastic
box on the optical table and the pipe shielding the
volume of air near the  laser beam. For a room temperature of 21
$^0C$, an air temperature change of $1 ~K$
will result in a 0.9 ppm change of air refractive index.
For a temperature variation of $0.5 ~mK$
in the pipe, $N_{meas} = 1200$ distance measurements, 
the estimated error will be $\sigma_R/R \sim 0.9
~ppm/K \times 0.5 ~mK \times \Omega^* \sim 42 ~ppb$, 
where the magnification factor $\Omega^* = 94$.

The total error from the above sources, when added in quadrature,
is $ \sim 89~ppb$, with the major error sources arising from the uncertainty of
fringe determination and the thermal drift. The estimated relative
error agrees well with measured relative spreads of $91~ppb$ in real data
for measured distance of about 0.45 meters.

Besides the above error sources, other sources can contribute to systematic 
bias in the absolute differential distance measurement. The major systematic 
bias comes from the uncertainty in the FSR of the F-P used to determine the 
scanned frequency range. The relative error would be
$\sigma_R/R \sim 50~ppb$ if the FSR were calibrated by a wavemeter with
a precision of $50 ~ppb$. A wavemeter of this precision was not available
for the measurements described here. The systematic bias from the multiple-distance-measurement
technique was also estimated by changing the starting point of the measurement window, 
the window size and the number of measurements, the uncertainties typically 
range from 10 to 50 nanometers. Systematic bias from uncertainties in 
temperature, air humidity and barometric pressure scales are estimated to be negligible.

\section{Conclusion}

An optical fiber FSI system was constructed to 
make high-precision absolute distance and vibration measurements. 
A design of the optical fiber with GRIN lens was 
presented which improves the geometrical efficiency significantly. 
Two new multiple-distance-measurement analysis techniques were presented to
improve distance precision and to extract the amplitude
and frequency of vibrations. Absolute distance measurement precisions of 
approximately 50 nm for distances 
ranging from 10 cm to 70 cm under laboratory conditions were achieved 
using the first analysis technique. 
The second analysis technique measures vibration frequencies ranging from 
0.1 Hz to 100 Hz with minimal amplitude of a few nanometers.
We verified an expected dispersion effect and confirmed its importance when
dispersive elements are placed in the interferometer.
Major error sources were estimated, and the observed errors were found to 
be in good agreement with expectation.

%\begin{acknowledgments}

This work is supported by the National Science Foundation and the Department of Energy of
the United States.

%\end{acknowledgments}

\bigskip
$^*$ Corresponding author, e-mail address: yhj@umich.edu

\end{document}